\documentclass[pra,letterpaper,twocolumn,aps,footinbib,superscriptaddress,showpacs,babel,longbibliography]{revtex4-1}

\usepackage{amsmath, upgreek, amssymb, graphicx, subfigure, paralist, dsfont, transparent}
\usepackage[colorlinks, linkcolor=blue, citecolor=blue, urlcolor=blue, breaklinks=true]{hyperref}
\usepackage{xspace}
\usepackage{siunitx}

\newcommand{\ket}[1]{|#1\rangle}
\newcommand{\bra}[1]{\langle #1|}
\newcommand{\braket}[1]{\langle #1\rangle}
\usepackage{subfigure}
\usepackage{textcomp}

\newcommand{\zpfz}{\ell_z}
\newcommand{\zpfr}{\ell_r}

\def\ppbar{\mbox{(anti-)}proton\xspace}
\def\ppbars{\mbox{(anti-)}protons\xspace}

\usepackage{color}
\definecolor{magnetron}{rgb}{1,0.647,0.0}
\definecolor{zyklotron}{rgb}{0.0,0.502,0.0}
\definecolor{redside}{rgb}{1,0.00,0.00}
\definecolor{blueside}{rgb}{0.122,0.467,0.706}
\begin{document}

\title{Elementary laser-less quantum logic operations with \ppbars in Penning traps}

\author{Diana Nitzschke}
\affiliation{Institut f\"{u}r Theoretische Physik und Institut f\"{u}r Gravitationsphysik (Albert-Einstein-Institut), Leibniz Universit\"{a}t Hannover, Appelstra\ss e 2, 30167 Hannover, Germany}

\author{Marius Schulte}
\affiliation{Institut f\"{u}r Theoretische Physik und Institut f\"{u}r Gravitationsphysik (Albert-Einstein-Institut), Leibniz Universit\"{a}t Hannover, Appelstra\ss e 2, 30167 Hannover, Germany}

\author{Malte Niemann}
\affiliation{Institut f\"{u}r Quantenoptik, Leibniz Universit\"{a}t Hannover,  Welfengarten 1, 30167 Hannover, Germany}

\author{Juan M.\ Cornejo}
\affiliation{Institut f\"{u}r Quantenoptik, Leibniz Universit\"{a}t Hannover,  Welfengarten 1, 30167 Hannover, Germany}

\author{Stefan Ulmer}
\affiliation{RIKEN, Ulmer Fundamental Symmetries Laboratory, 2-1 Hirosawa, 351-0198 Saitama, Japan}

\author{Ralf Lehnert}
\affiliation{Institut f\"{u}r Quantenoptik, Leibniz Universit\"{a}t Hannover,  Welfengarten 1, 30167 Hannover, Germany}
\affiliation{Indiana University Center for Spacetime Symmetries, Bloomington, IN 47405, U.S.A.}

\author{Christian Ospelkaus}
\email[]{christian.ospelkaus@iqo.uni-hannover.de}
\affiliation{Institut f\"{u}r Quantenoptik, Leibniz Universit\"{a}t Hannover, Welfengarten 1, 30167 Hannover, Germany}
\affiliation{Physikalisch-Technische Bundesanstalt, Bundesallee 100, 38116 Braunschweig, Germany}

\author{Klemens Hammerer}
\affiliation{Institut f\"{u}r Theoretische Physik und Institut f\"{u}r Gravitationsphysik (Albert-Einstein-Institut), Leibniz Universit\"{a}t Hannover, Appelstra\ss e 2, 30167 Hannover, Germany}

\begin{abstract}
Static magnetic field gradients superimposed on the electromagnetic trapping potential of a Penning trap can be used to implement laser-less spin-motion couplings that allow the realization of elementary quantum logic operations in the radio-frequency regime. An important scenario of practical interest is the application to $g$-factor measurements with single \ppbars to test the fundamental charge, parity, time reversal (CPT) invariance as pursued in the BASE collaboration~\cite{smorra_base_2015,smorra_parts-per-billion_2017, schneider_double-trap_2017}. We discuss the classical and quantum behavior of a charged particle in a Penning trap with a superimposed magnetic field gradient. Using analytic and numerical calculations, we find that it is possible to carry out a SWAP gate between the spin and the motional qubit of a single \ppbar with high fidelity, provided the particle has been initialized in the motional ground state. We discuss the implications of our findings for the realization of quantum logic spectroscopy in this system. 
\end{abstract}

\date\today

\maketitle

\section{Introduction}
Laser-based state manipulation and readout is the standard approach for cooling, state engineering and state readout for trapped ions in radio-frequency Paul traps with important applications in quantum information processing and metrology. An important mechanism is the implementation and application of the Jaynes-Cummings model in ion traps, which involves a coherent coupling between harmonic-oscillator motional and internal degrees of freedom of the atom. Its application is ubiquitous for ground state cooling, quantum logic gates and motional state engineering. 

Recent years have seen renewed interest~\cite{mintert_ion-trap_2001,ospelkaus_trapped-ion_2008} in the use of radio-frequency and microwave fields for this purpose because of the potentially better control compared to laser beams and because of the possibility to integrate the generation of control fields into scalable trap structures. Another reason to consider laser-less radio-frequency control fields for this purpose is that many systems of physical interest do not possess any reachable optical transitions for implementing this type of dynamics. A particularly challenging example is the case of a single \ppbar~\cite{smorra_parts-per-billion_2017}, which does not possess any electronic structure at all. 

Already in 1990, Heinzen and Wineland proposed a protocol~\cite{heinzen_quantum-limited_1990} that would enable full control over such a sub-atomic particle by coupling it to a laser-cooled atomic ion for $g$-factor measurements. The same ideas have later been applied to the Al$^+$ ion in the context of frequency metrology and are now known as quantum logic spectroscopy. These protocols rely on the Coulomb interaction between the particle of interest and the laser-cooled `logic' ion. While these protocols do shift a significant part of the control challenge to the atomic ion, for internal-state readout of the particle of interest, at least a SWAP gate between its internal and motional degrees of freedom needs to be carried out. In the case of the single-ion Al$^+$ clock~\cite{schmidt_spectroscopy_2005,rosenband_observation_2007}, this operation is realized using the $^3P_1$ laser pulse. 

In the case of the \ppbar, Heinzen and Wineland discussed the application of an oscillating magnetic field amplitude gradient for this purpose. These ideas have later inspired oscillating near-field entangling gates with trapped ions~\cite{ospelkaus_trapped-ion_2008}. When applied to the \ppbar, the implementation is extremely challenging because of the much smaller magnetic moment and because of the typically much bigger trap structures, which lead to orders of magnitude smaller oscillating near-field gradients. A viable alternative is the use of a static magnetic field gradient for this purpose. Penning traps can support very large static magnetic-field inhomogeneities~\cite{Ulmer2011}; in the case of $g$-factor measurements, one typically considers a very strong magnetic-field curvature induced by an embedded piece of ferromagnetic material to make the axial frequency depend on the internal state of the particle for spin-state readout via the continuous Stern-Gerlach effect~\cite{dehmelt_continuous_1986}. 

Here we propose to employ first-order (gradient) static magnetic-field inhomogeneities in Penning traps in order to implement spin-motional couplings as discussed by Mintert and Wunderlich in the context of radio-frequency Paul traps and quantum information processing~\cite{mintert_ion-trap_2001}. The gradients that can be generated by embedded ferromagnetic materials are typically much stronger than the oscillating near-field gradients that can be realized in a comparable scenario, making this approach our method of choice for implementing quantum logic spectroscopy of \ppbars. We analyze the classical and quantum behavior of a Penning trap with a superimposed static magnetic field gradient. Through analytical calculations and using numerical simulations, we find regimes where a SWAP gate can be carried out between the internal and motional degrees of freedoms of single \ppbars. Our findings are not limited to this case, but may be of more general use for the implementation of quantum logic spectroscopy and elementary quantum logic operations in Penning traps. 


\section{Penning trap with longitudinal gradient and transverse oscillating fields}
\label{QTrap}

\subsection{Conventional Penning trap}\label{sec:Title}
We first recapitulate the known quantum mechanical description of a particle in a conventional Penning trap~\cite{Brown_geonium_1986, Crimin_quantum_2017}. A quadrupole electric field confines the particle along the $z$ direction. The potential $V(\vec{r})$ and electric field $\vec{E}(\vec{r})$ can be expressed as
\begin{align}
V(\vec{r})&=V_{\mathrm{R}} C_2 (z^2-\frac{x^2}{2}-\frac{y^2}{2}), \nonumber \\
\vec{E}(\vec{r})&= V_{\mathrm{R}} C_2 \begin{pmatrix}
x\\
y\\
-2z
\end{pmatrix}.\label{5.6}
\end{align}
The field is produced by applying a voltage $V_{\mathrm{R}}$ to a set of typically cylindrical electrodes with their axis also aligned along the $z$ direction. For axial confinement the sign of the voltage needs to agree with the sign of the charge $q$. The parameter $C_2$ characterizes the geometry of the trap, where $\sqrt{1/C_2}$ is a trap specific length~\cite{smorra_base_2015}.
Confinement in the $x-y$ plane is achieved by means of a constant magnetic field along the $z$ axis with vector potential $\vec{A}(\vec{r})$ and magnetic field strength $\vec{B}(\vec{r})$,
\begin{align}\label{eq:B0}
\vec{A}_0(\vec{r})&=\frac{B_0}{2}\begin{pmatrix}
y\\
\:\:-x\\
\:\:0
\end{pmatrix}
&
\vec{B}_0(\vec{r})&=B_0\begin{pmatrix}
0\\
0\\
-1
\end{pmatrix}.
\end{align}

Note that here we align $\vec{B}_0$ in negative $z$ direction, as required for the antiproton (with $q=-e_0$). For the proton this direction should be reversed along with the directions of the other magnetic fields; $\vec{B}_1, \vec{B}_2$; introduced below. Making the replacements $q \rightarrow -q, B_0 \rightarrow -B_0, b \rightarrow -b$ and $B_2 \rightarrow -B_2$ then gives the identical Hamiltonian for the proton as compared to the antiproton considered here (note that $q \rightarrow -q$ also causes $\vec{\mu} \rightarrow -\vec{\mu}$ consequently).
The minimal coupling Hamiltonian for the motion of a particle with charge $q$ and mass $m$ in these fields is
\begin{align*}
H_\mathrm{mot}=\dfrac{(\vec{p}-q \vec{A}_0)^2}{2m}+ q V,
\end{align*}
where $\vec{p}$ is the momentum operator. We have $[r_k,p_l]=i\hbar\delta_{kl}$ for $k,l=x,y,z$. 
This Hamiltonian can be diagonalized~\cite{Brown_geonium_1986} and decomposes into three terms corresponding to independent harmonic oscillators, one for the motion along $z$ and two for the motion in the transverse direction
\begin{multline}\label{eq:Hmot}
H_\mathrm{mot}=\hbar\omega_z\left(a_z^\dagger a_z+\frac{1}{2}\right)\\
+\hbar\omega_+\left(a_c^\dagger a_c+\frac{1}{2}\right)
-\hbar\omega_-\left(a_m^\dagger a_m+\frac{1}{2}\right).
\end{multline}
Here we have defined the axial, modified cyclotron, and magnetron frequency,
\begin{align}\label{eq:frequencies}
\omega_z&=\sqrt{2 V_{\mathrm{R}} C_2 \frac{q}{m}},&
\omega_+&=\frac{\omega_c}{2}+\Omega_c,&
\omega_-&=\frac{\omega_c}{2}-\Omega_c,
\end{align}
where $\omega_c= -q B_0/m$ is the cyclotron frequency and $\Omega_c>0$ is defined by
$
\Omega_c^2={\omega_c^2}/4-{\omega_z^2}/2.    
$
For common Penning trap parameters these frequencies obey the hierarchy $\omega_+\gg\omega_z\gg\omega_-$. The annihilation operators for $k=x,y,z$ are 
\begin{align}\label{eq:annih}
a_k&=\dfrac{1}{\sqrt{2}}\left(\dfrac{1}{\ell_k}r_k+\frac{i\ell_k}{\hbar}p_k\right),
\end{align}
where $\ell_z$ and $\ell_x=\ell_y\equiv\ell_r$ are the characteristic length scales of the harmonic oscillators for axial and radial motion, $\zpfz=\sqrt{\hbar/m\omega_z}$ and $\zpfr=\sqrt{\hbar/m\Omega_c}$. Finally, annihilation operators for cyclotron and magnetron motion  are $a_c =(a_x+ia_y)/\sqrt{2}$ and $a_m =(a_x-ia_y)/\sqrt{2}$. We have $[a_k,a_l^\dagger ]=\delta_{kl}$ for $k,l=z,c,m$.
%

%
For a spin-$1/2$ particle with magnetic moment $\vec{\mu}$ and gyromagnetic factor $g$ the magnetic dipole energy is
\begin{align}\label{eq:Hspin}
H_\mathrm{spin}&=-\vec{\mu}\cdot\vec{B}_0=\frac{\hbar}{2}\omega_{\mathrm{L}}\sigma_z,&
\omega_{\mathrm{L}}&=\frac{g}{2}\omega_c.
\end{align}
The total Hamiltonian for a conventional Penning trap configuration is
\begin{align*}
H_0=H_\mathrm{mot}+H_\mathrm{spin},
\end{align*}
where $H_\mathrm{mot}$ and $H_\mathrm{spin}$ are given in Eqs.~\eqref{eq:Hmot} and \eqref{eq:Hspin}.


\subsection{Gradient field}\label{sec:Title2}

Next, we will include an additional magnetic field providing a constant field gradient along $z$. We describe the gradient field by a vector potential $\vec{A}_1(\vec{r})$ with the corresponding magnetic field strength $\vec{B}_1(\vec{r})$,
\begin{align}\label{eq:gradient}
    \vec{A}_1(\vec{r})&=\frac{b}{2}\begin{pmatrix}
	\:\:zy\\
	-zx\\
	\:\:0
	\end{pmatrix},
	&
	\vec{B}_1(\vec{r})&=b\begin{pmatrix}
	x/2\\
	y/2\\
	-z
	\end{pmatrix}.
\end{align}
The parameter $b$ describes the magnitude of the gradient. The complete Hamiltonian is given by
\begin{align} \label{eq:H0and1}
H&=\dfrac{\big(\vec{p}- q (\vec{A}_0+\vec{A}_1)\big)^2}{2m}-\vec{\mu}\cdot(\vec{B}_0+\vec{B}_1)+ q V \nonumber \\
&=H_0+H_1,
\end{align}
where $H_1$ collects all terms added by the gradient field. It will be useful to characterize the strength of the gradient by a dimensionless parameter
\begin{align}\label{eq:epsilon}
\epsilon=\frac{b\zpfz}{2 \sqrt{2} B_0},
\end{align}
which measures the relative change of the magnetic field in units of the zero point fluctuations $\zpfz$ of the ground state of motion along $z$.

In terms of the creation and annihilation operators introduced in the previous section $H_1$ can be written as (with $2\Omega_c \approx \omega_c$ and up to constant terms)
\begin{align}\label{eq:H1}
H_1&=\hbar\omega_c\epsilon
\left(a_z+a_z^\dagger\right)
\left(\frac{g}{2}\sigma_z+1+2a_c^\dagger a_c+a_ca_m+a_c^\dagger a_m^\dagger\right)\nonumber\\
&\quad-\hbar\omega_c\frac{g\epsilon}{2}\sqrt{\frac{\omega_z}{\omega_c}}\left(\sigma^+\left(a_m+a_c^\dagger\right)+\sigma^-\left(a_c+a_m^\dagger\right)\right)\nonumber\\
&\quad+\hbar\omega_c\epsilon^2\left(a_z+a_z^\dagger\right)^2(a_c^\dagger +a_m)(a_m^\dagger+a_c).
\end{align}
We note that the Hamiltonian $H$ commutes with the $z$~component of the total angular momentum, which we define in dimensionless form as
\begin{align*}
J_z&=(L_z+S_z)/\hbar=(xp_y-yp_x)/\hbar+\sigma_z/2\\
&=a_m^\dagger a_m-a_c^\dagger a_c +{\sigma_z}/{2}.
\end{align*}
The desired coupling among spin and motion will be attained from the first, Stern-Gerlach-like term proportional to $\left(a_z+a_z^\dagger\right)\sigma_z$ in Eq.~\eqref{eq:H1}.

\subsection{Transverse oscillating field}\label{sec:Title3}

In order to produce a resonant coupling among the axial mode and the spin we add to the previous configuration a transverse oscillating magnetic field $\vec{B}_2(\vec{r},t)$ with a frequency close to the first axial sideband on the spin transition, such that $\omega\approx\omega_\text{L}\pm\omega_z$. In combination with the constant magnetic field in the axial direction the transverse oscillating field can produce Rabi cycles. What will be shown in this section is that if the oscillating field is applied with the right frequency these Rabi cycles actually correspond to sideband transitions involving the axial mode. The transverse oscillating field is given by the vector potential and magnetic field strength 
\begin{align}
\vec{A}_{2}(\vec{r},t)&=\frac{B_2}{2}\left(\begin{array}{c} -z\sin{(\omega t)} \\ z\cos{(\omega t)} \\ x\sin{(\omega t)} - y\cos{(\omega t)}\end{array}\right), 
\\
\vec{B}_{2}(t)&=B_2\left(\begin{array}{c} -\cos{(\omega t)} \\ -\sin{(\omega t)} \\0 \end{array}\right).
\end{align}
The full Hamiltonian is
\begin{align*}
H&=\dfrac{\big(\vec{p}-q (\vec{A}_0+\vec{A}_1+\vec{A}_2)\big)^2}{2m}+q V-\vec{\mu}\cdot\!(\vec{B}_0+\vec{B}_1+\vec{B}_2)\\
&=H_0+H_1+H_2,
\end{align*}
where we collect all terms added by the transverse field in $H_2$. The explicit form of $H_2$ is given in Appendix~\ref{sec:dis}. We show there that the only relevant term in $H_2$ is
\begin{align*}
H_2\simeq-\vec{\mu}\cdot\vec{B_2}=\frac{\hbar\Omega}{2}(\sigma^+e^{-i\omega t}+\sigma^-e^{i\omega t}),
\end{align*}
where the Rabi frequency is $\Omega={-q g B_2}/2m$. All other terms are either small, non-resonant or both.

In order to see that the Stern-Gerlach term in $H_1$ and the spin flips in $H_2$ together can give rise to resonant sideband transitions, it is useful to apply a unitary transformation which absorbs the Stern-Gerlach term~\cite{mintert_ion-trap_2001}
\begin{align*}
\tilde{H}&=e^{S}He^{-S}=\tilde{H}_0+\tilde{H}_1+\tilde{H}_2,\\
S&=(\eta J_z+\alpha)(a_z^\dagger-a_z).
\end{align*}
The dimensionless parameter $\eta$ is chosen such that the Stern-Gerlach term from $H_1$ is canceled in $\tilde{H}_1$. Furthermore, $\alpha$ is adapted in order to remove any mean force on the particle in $z$ direction. These conditions yield
\begin{align}\label{eq:eta}
    \eta&=\frac{\epsilon g\omega_c}{\omega_z}, & \alpha&=\frac{\omega_c\epsilon}{\omega_z}.
\end{align}
For Hamiltonian $\tilde{H}_2$ one finds
\begin{align*}
\tilde{H}_2=\frac{\hbar\Omega}{2}\big(\sigma^+ e^{\eta (a_z^\dagger -a_z)}e^{-i\omega t}+\sigma^- e^{-\eta (a_z^\dagger -a_z)}e^{i\omega t}\big).
\end{align*}
In this picture it is evident that $\eta$ is an effective Lamb-Dicke factor setting the strength of sideband transitions. Assuming $\eta\ll 1$, we perform a Lamb-Dicke expansion to first order,
\begin{align}
\tilde{H}_2&\simeq\frac{\hbar\Omega}{2}\big(\sigma^+ e^{-i\omega t}+\sigma^- e^{i\omega t}\big)\nonumber\\
&\quad+\frac{\hbar\Omega\eta}{2}\big(\sigma^+ e^{-i\omega t}-\sigma^- e^{i\omega t}\big)\big(a_z^\dagger-a_z\big).
\end{align}
It is straight forward to derive the transformed Hamiltonians $\tilde{H}_0$ and $\tilde{H}_1$. Both of these terms have at most a linear dependence on $\eta$, such that the Lamb-Dicke approximation does not change their structure. We refrain from giving their explicit form here.

In this picture the Hamiltonian still has an explicit time dependence via $\tilde{H}_2$. We remove the time dependence by changing to a frame rotating with the frequency of the transverse field,
\begin{align}
\bar{H}&=e^{i\omega J_z t}\tilde{H}e^{-i\omega J_z t}-\hbar\omega J_z.
\end{align}
Defining the detuning of the transverse oscillating field from the effective spin transition frequency,
\begin{align}\label{eq:detuning}
\Delta=\omega_{\mathrm{L}}-2g\epsilon\alpha\omega_c-\omega,
\end{align}
the final Hamiltonian can be expressed as
\begin{align}\label{eq:Htot}
\bar{H}&=\dfrac{\hbar\Delta}{2}\sigma_z+\hbar\omega_za_z^\dagger a_z
+\frac{\hbar\Omega}{2}(\sigma^++\sigma^-) \nonumber \\
&\quad+\frac{\hbar\Omega\eta}{2}(\sigma^+ -\sigma^-)(a_z^\dagger-a_z)+\bar{H}_\mathrm{rest}.
\end{align}
The third term on the right hand side describes carrier transitions of the spin at a Rabi frequency $\Omega$. The fourth term describes the desired coupling of spin and motion along $z$ via sideband transitions, adding or removing motional quanta along with spin flips at an effective Rabi frequency $\Omega\eta$. In $\bar{H}_\mathrm{rest}$, we collect all terms that either do not couple to spin and motion in $z$ direction or do so only in order $\epsilon$ (defined in Eq.~\eqref{eq:epsilon}) or higher,
\begin{align}\label{eq:Hrest}
\bar{H}&_\mathrm{rest}=\hbar(\omega_++\omega) a_c^\dagger a_c-\hbar(\omega_-+\omega) a_m^\dagger a_m+\hbar\omega_z\eta^2J_z^2\nonumber\nonumber\\
&-\hbar\omega_c\epsilon
\Big\{g\eta J_z\sigma_z+\dfrac{g}{2}\sqrt{\dfrac{\omega_z}{\omega_c}}\big(\sigma^-(a_c+a_m^\dagger)+\mathrm{h.c.}\big)\nonumber\\
&\quad-\big(a_z+a_z^\dagger-2\eta J_z-2\alpha\big)\big(2a_c^\dagger a_c+a_ca_m+a_c^\dagger a_m^\dagger\big)\nonumber\\
&\quad+g\big(a_m^\dagger a_m-a_c^\dagger a_c\big)\big(a_z^\dagger+a_z\big)\Big\}\nonumber\\
&+\hbar\omega_c\epsilon^2\big(a_z+a_z^\dagger-2\eta J_z-2\alpha)^2(a_c^\dagger +a_m)(a_m^\dagger+a_c).
\end{align}

The challenge is now to identify a parameter regime where the sideband transitions in the Hamiltonian \eqref{eq:Htot} can be exploited for mapping a spin excitation to $z$~motion while suppressing the undesired coupling of the spin to cyclotron and magnetron motion implied by Hamiltonian \eqref{eq:Hrest}. Since these processes happen at a rate $\omega_c\epsilon$, and sideband transitions happen at rate $\Omega\eta$ where $\eta\propto\epsilon$, cf.\ Eq.~\eqref{eq:eta}, it is clear that this requires to make the right trade-off.

\section{Numerical Case Study}
\label{NTrap}

\begin{center}
\begin{table*}[bt]
\begin{tabular}{|c|c c c c c|} 
 \hline
 & Parameter & Symbol & Value & Unit & Eq.\\ [0.5ex] 
 \hline\hline
 Independent trap parameters & Longitudinal magnetic field & $B_0$ & 3.00 & \si{\tesla} & \eqref{eq:B0}\\
 & Longitudinal trap frequency & $\omega_z$ & 91.49 & $2\pi\,\si{kHz}$ & \eqref{eq:frequencies} \\
 & Magnetic field gradient & $b$ & 1200.00 & \si{\tesla/\meter} & \eqref{eq:gradient}\\
 [0.5ex] 
 \hline\hline
 Derived trap parameters & Cyclotron frequency & $\omega_c$ & 45745.13 & $2\pi\,\si{kHz}$ & \eqref{eq:frequencies} \\
 & Modified cyclotron frequency & $\omega_+$ & 45745.04 &  $2\pi\,\si{kHz}$ & \eqref{eq:frequencies}\\
 & Magnetron frequency & $\omega_-$ & 91.49 & $2\pi\,\si{Hz}$ & \eqref{eq:frequencies}\\
 & Relative field gradient per zero point fluctuation & $\epsilon$ & $4.68 \!\cdot\! 10^{-5}$ & - & \eqref{eq:epsilon} \\
 & Mean displacement in $z$ per zero point fluctuation & $\alpha$ & $2.34 \!\cdot\! 10^{-2}$ & - & \eqref{eq:eta} \\
 & Effective Lamb-Dicke parameter & $\eta$ & $1.31 \!\cdot\! 10^{-1}$ & - & \eqref{eq:eta} \\
 & Gradient induced shift of spin resonance & $2 g \epsilon \alpha \omega_c$ & 558.95 & $2\pi\,\si{Hz}$ & \eqref{eq:detuning}\\
  [0.5ex]
 \hline\hline
 Pulse parameters & Rabi frequency & $\Omega$ & 1.91 & $2\pi\,\si{kHz}$ & \eqref{eq:Htot}\\
 & Effective Rabi frequency for sideband transition & $\eta\Omega$ & 250.00 & $2\pi\,\si{Hz}$ & \eqref{eq:Htot}\\
 & Spurious coupling of spin to cyclotron/magnetron mode & $\epsilon\omega_c$ & 2.14 & $2\pi\,\si{\kHz}$ & \eqref{eq:Hrest}\\
 & Pulse duration for $\pi$-pulse & $\tau$ & 2.00 & \si{ms} & \eqref{eq:tau} \\
 & Suppression of carrier transition & $\Omega/\omega_z$ & $2 \!\cdot\! 10^{-2}$ & - & \\
  [0.5ex]
 \hline
\end{tabular}
\caption{Case study for sideband pulses coupling spin and motion of an \ppbar.}
\label{table:1}
\end{table*}
\end{center}

\begin{figure}[t]
	\centering
	\includegraphics[width=\linewidth]{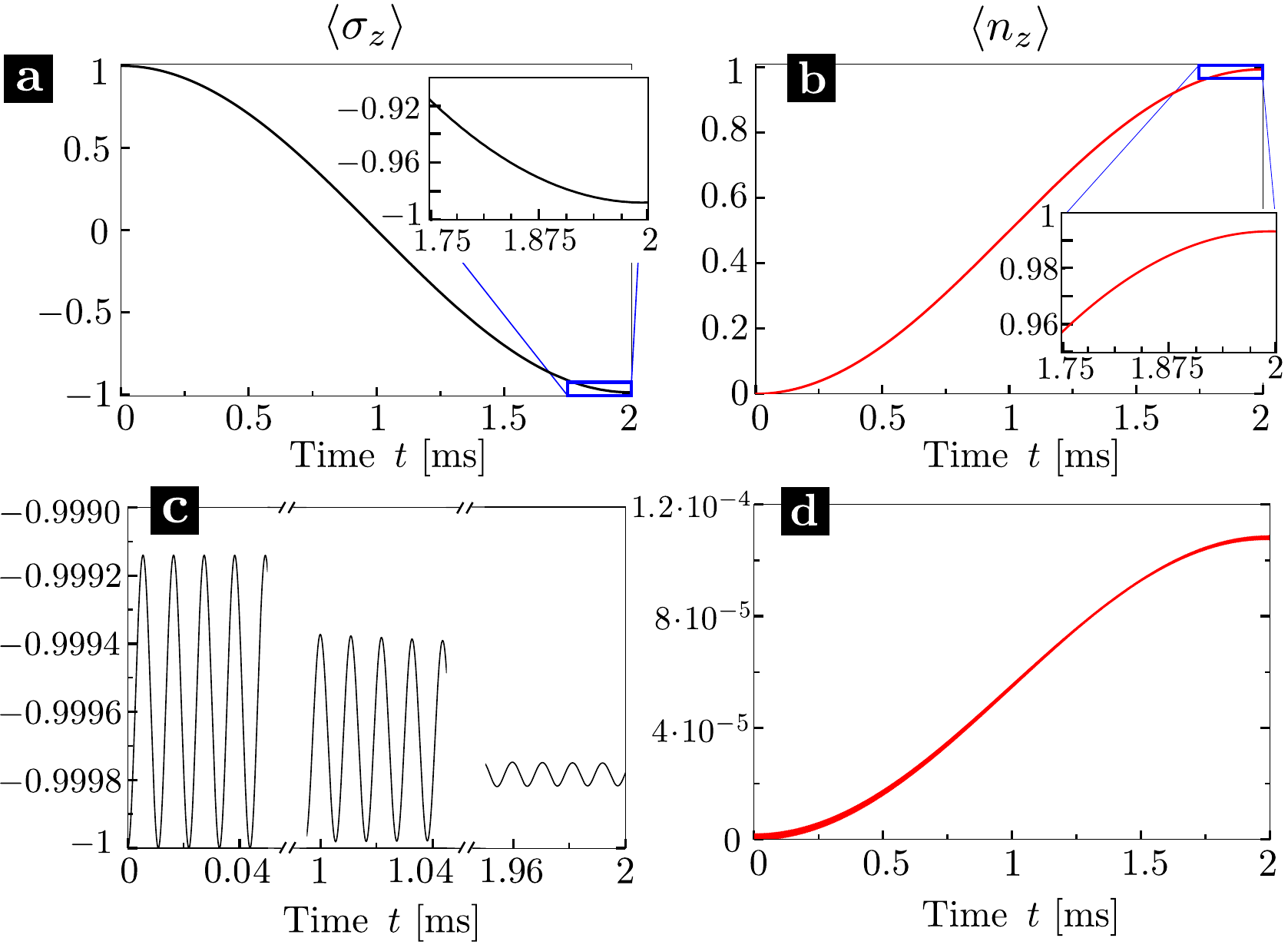}
	\caption{Time traces for the spin polarization $\langle \sigma_z \rangle$ (black) and average quanta in the axial mode $\braket{n_z}$ (red) when driving a red sideband, i.e. $\Delta = \omega_z$. At the start of the dynamics, all motional modes are in their ground state. We show the dynamics starting from $\ket{\uparrow} \otimes \ket{0}_z$ in parts a-b and $\ket{\downarrow}\otimes \ket{0}_z$ in parts c-d. The parameters are as shown in Table~\ref{table:1}.
	}
	\label{fig:red_sideband1}
\end{figure}

In this section we consider a trapped \ppbar 
and show that the sideband transitions introduced above can be implemented faithfully and with only small errors due to the perturbation terms. 
This analysis employs the numerical values for the antiproton
($q=-e_0=-1.6\cdot10^{-19}$\,C, $m_p=1.67\cdot10^{-27}$\,{kg},
$g = 5.5857$),
but our conclusions remain valid for the proton~\footnote{With the replacements mentioned in Sec.~\ref{QTrap}, all parameters remain at the same value and one would have $\omega_c = q B_0/m > 0$ and $\Omega = q g B_2/(2m) > 0$ for the proton as well.}. 
We consider first one particular set of parameters compatible with state-of-the-art Penning trap designs. Later on, we will study gate error probabilities for larger sets of parameters.

We imagine an \ppbar has been placed in a Penning trap with a field configuration as discussed in Sec.~\ref{QTrap}. With the parameters presented in Table~\ref{table:1} it is possible to exploit the sideband transitions coupling spin and motion along $z$, as described by Hamiltonian \eqref{eq:Htot}, without having significant contributions from the coupling to cyclotron and magnetron mode from the terms in Hamiltonian \eqref{eq:Hrest}. 

As a reference we briefly discuss the scenario of an ideal $\pi$ pulse on the red sideband: Assume the \ppbar is prepared in its ground state $\ket{0,0,0}$ of motion in $z$, cyclotron and magnetron modes and the spin state is $\ket{\uparrow\,}$ along $z$. This should be achievable through sympathetic cooling to the ground state on the axial mode \cite{goodwin_resolved-sideband_2016} and mode coupling between the radial and axial modes \cite{cornell_mode_1990}. Ideally, a pulse in the transverse field of duration 
\begin{equation}\label{eq:tau}
    \tau = \pi/\Omega\eta
\end{equation}
and oscillating at a detuning $\Delta=\omega_z$ from the effective spin resonance frequency will effectively convert the spin excitation into $z$ motion without affecting the other modes, $\ket{\uparrow\,}\otimes\ket{0,0,0}\rightarrow \ket{\downarrow\,}\otimes\ket{1,0,0}$. At the same time, if the spin was initially in state $\ket{\downarrow\,}$ no coupling to motion occurs, $\ket{\downarrow\,}\otimes\ket{0,0,0}\rightarrow \ket{\downarrow\,}\otimes\ket{0,0,0}$.
In this way any spin superposition will be mapped onto the state of motion,
\begin{multline*}
    \big(c_\uparrow\ket{\uparrow\,}+c_\downarrow\ket{\downarrow\,}\big)\otimes\ket{0,0,0}
    \rightarrow\\
    \ket{\downarrow\,}\otimes\big(c_\uparrow\ket{1,0,0}+c_\downarrow\ket{0,0,0}\big).
\end{multline*}
The transfer will of course work equally well for spin mixtures. The excitation in the $z$ mode of motion can subsequently be read out via, e.g., further quantum logic operations.

In Fig.~\ref{fig:red_sideband1} we show the result of a numerical solution of the Schr{\"o}dinger equation for the complete Hamiltonian \eqref{eq:Htot}, including even the terms of second order in the Lamb-Dicke parameter, that is $({\hbar \Omega\eta^2}/{4})  (a_z^\dagger-a_z)^2 (\sigma^+ +\sigma^-)$. We truncate the Hilbert space of motional modes at Fock state $5$, which is sufficient in this case as the entire dynamics is limited to the lowest Fock states only.\\ Figure~\ref{fig:red_sideband1} shows the spin polarization $\braket{\sigma_z(t)}$ and the average number of quanta $\braket{n_z(t)}$ along the $z$ direction versus time for the initial state $\ket{\uparrow\,}\otimes\ket{0,0,0}$ (in parts a and b) and $\ket{\downarrow\,}\otimes\ket{0,0,0}$ (in parts c and d), respectively. Figure~\ref{fig:red_sideband1}a and Figure~\ref{fig:red_sideband1}b clearly show the spin excitation oscillating over to the motional degree of freedom within a time $\pi/\eta\Omega=2\,\si{ms}$ for the pulse parameters given in Table~\ref{table:1}. Figure~\ref{fig:red_sideband1}c and Figure~\ref{fig:red_sideband1}d illustrate the effects of spurious dynamics due to coupling to cyclotron and magnetron motion on the order of $\epsilon$, as expected.

\begin{figure}[t]
	\centering
    \includegraphics[width=\columnwidth]{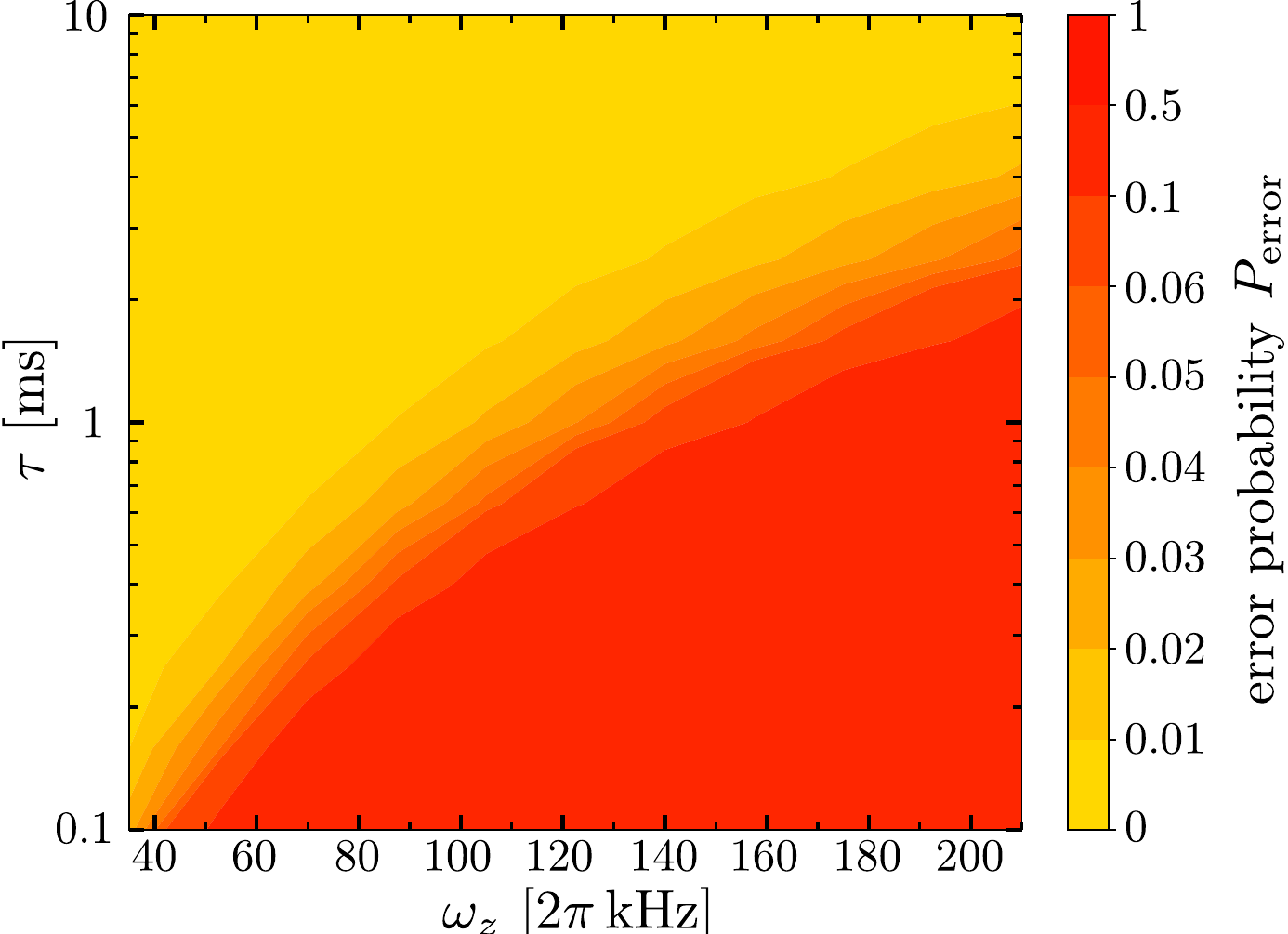}
    \caption{Error probability for a spin measurement based on the sideband SWAP as a function of the duration $\tau$ and axial frequency $\omega_z$. The non-linear color scale shows two major regimes of operation. In the bottom right, the red region signals unfeasible SWAP operations with errors beyond 50\%. In the top left, the yellow region corresponds to SWAP interactions with error probabilities at most 1\%. This should be viewed as an upper bound to the readout error as for this figure the numerical simulations considered the spin and axial motion only. We are thus not able to exclude the presence of additional small errors below 1\% resulting from the neglected cyclotron or magnetron mode.}\label{fig:errorprob}
\end{figure}

When the state swap from spin to motion is to be used as a spin measurement, we can quantify the intrinsic imperfections of the readout by studying the error probability.
Specifically, readout errors occur with probability $P(n_z=0\vert\uparrow)$ when starting with the spin in the excited state and with probability $P(n_z\neq0\vert\downarrow)$ when starting in the ground state. The two cases describe, respectively, the absence of state transfer from the excited spin state or the faulty measurement of an excitation in the $z$ mode by off-resonant driving. We define the total error probability as $P_\mathrm{error}=(P(n_z=0\vert\uparrow)+P(n_z\neq0\vert\downarrow))/2$, where an equal \textit{a priori} probability for both spin states was assumed. Figure~\ref{fig:errorprob} shows the total error probability versus pulse duration $\tau$ and longitudinal confinement $\omega_z$. The Rabi frequency is scaled such that $\eta\Omega\tau=\pi$, in order to assure a proper state swap in each case.

So far we restricted our study to the ideal case of an \ppbar with perfect ground state cooling of all motional modes. If we now add the effect of small thermal occupations, we find that the indirect spin measurement is robust against a single excitation of the axial mode, but sensitive to the cyclotron or magnetron mode on the level of single quanta.
Let us consider an initial state
\begin{equation}\label{eq:init_z}
    \rho_z=p_{z0}\ket{0}_z\bra{0}+p_{z1}\ket{1}_z\bra{1},
\end{equation}
with an excitation of the first Fock state with probability $p_{z1}=1-p_{z0}$. In this case it turns out that the error probability is independent of $p_{z1}$ and limited only by the intrinsic error of the sideband ($P_0 = 3.2 \!\cdot\! 10^{-3}$ for the parameters of Table~\ref{table:1}).
The insensitivity results from the fact that the sideband transition transfers a particle in the $\ket{\!\downarrow}$ from $\ket{1}_z \rightarrow \ket{0}_z$ on the one hand and a particle in the $\ket{\!\uparrow}$ state would only transition from $\ket{1}_z \rightarrow \ket{2}_z$ on the other hand. So in both cases the same (correct) measurement result is still obtained, leaving the probabilities $P(n_z=0\vert\uparrow)$, $P(n_z\neq0\vert\downarrow)$ and thus the total error invariant. Higher Fock states however will directly result in additional readout errors and their contributions should therefore be kept as small as possible.
Similarly we find that occupations of the cyclotron and magnetron mode, again with a single excitation as in Eq.~\eqref{eq:init_z} but with probabilities $p_{c1}$ and $p_{m1}$, give rise to significant additional errors.
For example, a single phonon in the cyclotron mode leads to a shift in the spin transition frequency by $ 4 \alpha \eta \omega_z $ due to the coupling term $-4\hbar\omega_c\epsilon\eta J_z a_c^\dagger a_c$ in Eq.~\eqref{eq:Hrest}. With the parameters of Table~\ref{table:1}, this additional detuning, $\sim 10^{-2} \omega_z$, significantly suppresses the sideband dynamics and results in a readout error $P_{\mathrm{error}, c} = P_0 + 0.483\, p_{c1}$. Note that the error close to $1/2$ is related to weighing the contributions of both spin states equally and for $\ket{\downarrow}$ remaining in the initial state will technically give the correct measurement result even if no swap process occurs.
Interestingly, for the magnetron mode a similar coupling term does not exist in Eq.~\eqref{eq:Hrest}. Nevertheless the unwanted terms therein give rise to a smaller, but considerable, readout error ($P_{\mathrm{error}, m} = P_0 + 0.017\, p_{m1}$ for the values of Table~\ref{table:1}).

\begin{figure}[t]
	\centering
    \includegraphics[width=\columnwidth]{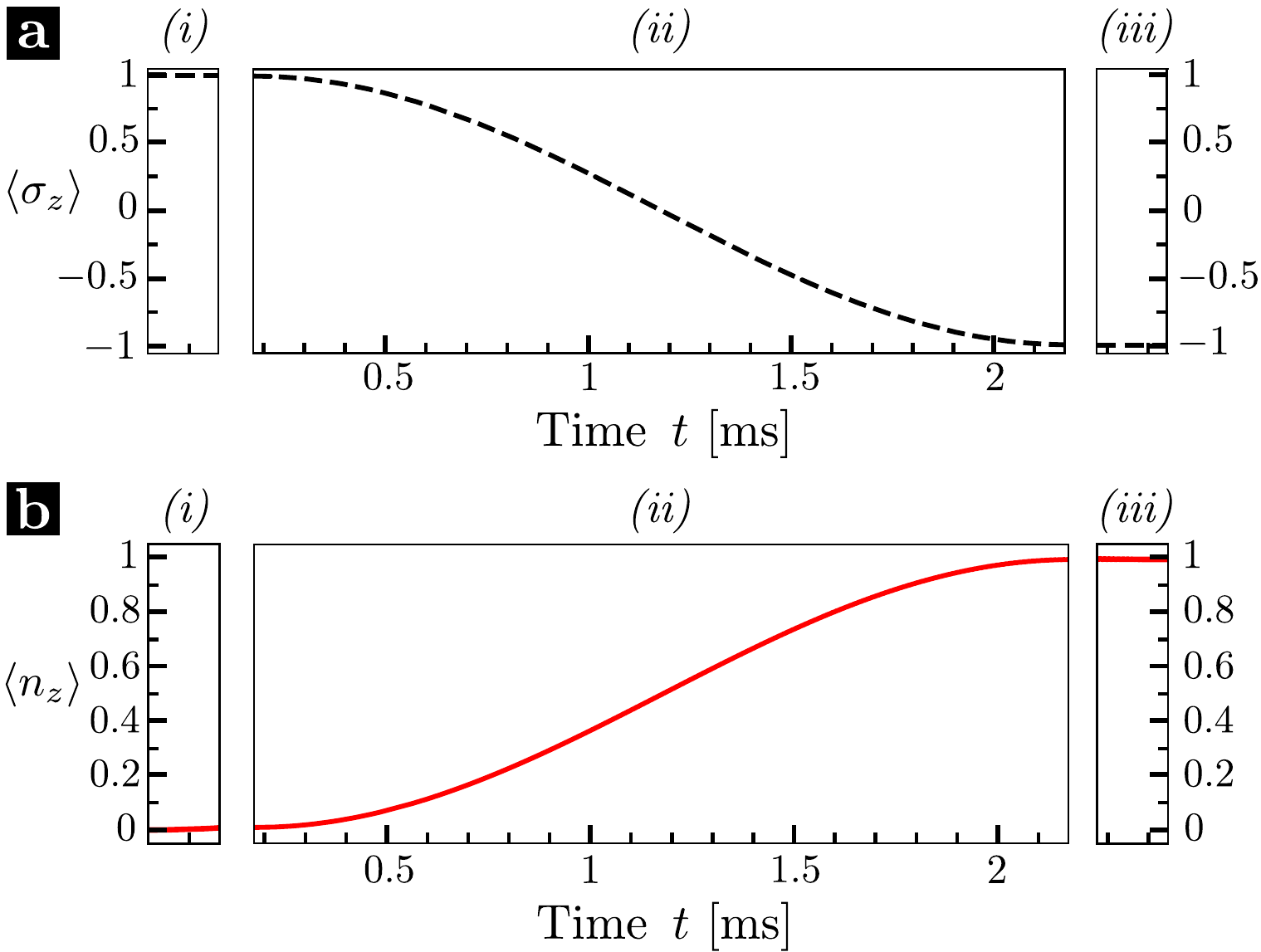}
    \caption{Time traces for transport and SWAP operations. (a) spin polarization and (b) average axial mode occupation during \textit{(i)} the initial splitting of the wavepackets, \textit{(ii)} the driven red sideband and \textit{(iii)} recombination.}
    \label{fig:3steps}
\end{figure}

One further simplification we made was to assume that the particle already starts at its equilibrium position within the large magnetic field gradient when the transverse oscillating field is switched on.
A more realistic approach to inducing the sideband transitions would be to study the transport of the particle from a trapping region with the homogeneous magnetic field only, as in the conventional Penning trap, into the magnetic field gradient.
A minimal model for this process can be constructed from the theoretical framework introduced above. The state swap can be divided into three steps: \textit{(i)} Transport of the particle into the field gradient with the transversal oscillating field turned off, i.e. $\Omega = 0$. In this stage the spin dependent splitting of the wave packet and the adjustment of the equilibrium position is described by the Hamiltonian in Eq.~\eqref{eq:H0and1} with a time dependent gradient $b(t)$ and correspondingly $\epsilon (t)$. We neglect terms of order $\epsilon^2$. We consider here the case of a linear increase, $\epsilon (t) = 4.68 \!\cdot\! 10^{-7} \omega_z t $, of the gradient up to $b=1200\,\mathrm{T}/\mathrm{m}$ at $t_1 = 100/\omega_z$ in accord with the above case study.
To avoid additional excitation of the axial mode we choose a slow increase, $ t_1 \gg 1/\omega_z $, over many oscillation periods. \textit{(ii)} Afterwards the oscillating field is switched on for duration $\tau$ to generate the swap between spin state and axial motion. Here the gradient is fixed. The dynamics in this step corresponds to the case study discussed above. \textit{(iii)} In the final step, the particle is being transported out of the gradient in order to perform the subsequent measurement of the axial mode. We reduce the gradient back to zero so that $\epsilon(t)= \epsilon - 4.68 \!\cdot\! 10^{-7} \omega_z t$.
The overall results for spin polarization and mean excitation of the axial mode are shown in Fig.~\ref{fig:3steps}. We find that there are no additional readout errors from this simple transport model.

\section{Newtonian treatment of Penning trap with gradient field}
\label{classical}

The previous section 
contains a fully  quantum-mechanical treatment of a Penning-trap setup 
with a magnetic gradient field 
permitting numerical investigations 
of the quantum dynamics in such a field configuration.
The goal of this section 
is to complement our numerical calculations 
with basic analytical results.
We focus on the effect of the gradient field $\vec{B}_1$ 
on the trap frequencies:
these are measurable key parameters characterizing the system,
and their study is standard 
in both the conventional homogeneous $\vec{B}$-field 
and magnetic-bottle configurations.
We will work classically at the  Newtonian level. 
An analytical quantum-mechanical treatment,
which would be desirable for the description of more complex physical phenomena, 
such as spin-flip transitions,
lies outside the scope of this section.

The nonrelativistic classical motion
of a point charge $q\neq 0$ with mass $m>0$ 
in a Penning trap 
supplemented by a magnetic-field gradient
is governed by
\begin{equation}
\label{fullEoM}
\ddot{\vec{r}}= \tfrac{1}{2}\omega_z^2
\big[\vec{r}-3(\vec{r}\cdot\hat{z})\hat{z}\big]
-\dot{\vec{r}}\times\! \Big
(\omega_c\hat{z}+\frac{q b}{2m}
\big[\vec{r}-3(\vec{r}\cdot\hat{z})\hat{z}\big]\Big),
\end{equation}
where we have employed 
the usual definitions of
\begin{equation}
\label{frequencyDefs}
\omega_z^2 \equiv 2 V_{\mathrm{R}} C_2 \frac{q}{m}\,,\qquad
\omega_c  \equiv -\frac{q B_0}{m}\,,
\end{equation}
as introduced in Sec.~\ref{sec:Title}.
In this section, 
we require the ordinary condition $q V_{\mathrm{R}}>0$ 
for axial confinement. 
We also take $\omega_c>0$,
a choice that 
requires selecting the $z$ axis 
antiparallel to $q \vec{B}_0$~\cite{Brown_geonium_1986}. 
The quantities 
\begin{equation}
\label{ECons}
E_0=\tfrac{1}{2}m\dot{\vec{r}}^{\,2}-\tfrac{1}{4}m\omega_z^2
\big[
\vec{r}^{\,2}-3(\vec{r}\cdot\hat{z})^2
\big]
\end{equation}
and 
\begin{equation}
\label{Lz}
L_z=
m\rho^2\dot{\phi}-\tfrac{1}{2} (m \omega_c - q b z )\rho^2\,,
\end{equation}
where $(\rho,\phi,z)$ are cylindrical coordinates,
are conserved 
due to time-translation invariance 
and symmetry under rotations about the $z$ axis.

As opposed to the homogeneous $\vec{B}$-field case,
Eq.~(\ref{fullEoM}) is no longer linear, 
and the axial and radial motion no longer decouple.
However,
proceeding in cylindrical coordinates
and employing angular-momentum conservation~(\ref{Lz}), 
$\dot{\phi}$ can be eliminated 
from the $\hat{\rho}$ and $\hat{z}$ components 
of Eq.~(\ref{fullEoM}):
\begin{align}
\label{rho_z_EoM}
m\ddot{\rho}&= F_\rho(\rho,z)\,,\nonumber\\
m\ddot{z}&= F_z(\rho,z)\,,
\end{align}
where $F_\rho(\rho,z)$ and $F_z(\rho,z)$ 
are the effective forces in $\hat{\rho}$ and $\hat{z}$ direction,
respectively.
Explicit expressions for these forces
can be read off directly from Eq.~(\ref{fullEoM}).
With this decoupling,
we may proceed by 
solving the system~(\ref{rho_z_EoM})
followed by 
determining $\phi(t)$ via Eq.~(\ref{Lz}). 

We begin by transforming to the following quantities
to expose the scaling behavior 
and the independent parameters of the system~(\ref{rho_z_EoM}):
\begin{align}
\label{Defs}
\tilde{\rho}\equiv&\;\frac{\rho}{\zpfz}\,,
& \tilde{z}\equiv&\;\frac{z}{\zpfz}\,,
&l_z\equiv&\;\xi\frac{L_z}{\hbar}\,,
& \zeta\equiv&\;\sqrt{2}\,\frac{\omega_z}{\omega_c}\,.
\end{align}
Here, 
$\zpfz$ is the harmonic oscillator length scale introduced in Eq.~\eqref{eq:annih}. 
The dimensionless radial and axial position variables
$\tilde{\rho}$ and $\tilde{z}$ 
are measured in units of $\zpfz$.
Note that $\tilde{\rho}$ should be non-negative,
whereas $\tilde{z}$ may exhibit both signs.
In the definition of $l_z$, 
$\xi \equiv L_z/|L_z|$ denotes the sign of $L_z$.
Thus, 
$l_z$ is dimensionless and non-negative; 
reducing $l_z$ to double-digit values 
signals the growing 
importance of quantum effects.
In the case of an ordinary Penning trap,
$0<\zeta<1$ must hold,
and this condition remains essential 
for our perturbative approach below.  

With these definitions,
we find 
\begin{align}
\label{eff-forces}
F_\rho(\tilde{\rho},\tilde{z})=&\,
\frac{m \omega_z^2 \zpfz }
{\zeta^2\tilde{\rho}^3}
\big[l_z^2 \zeta^2
-2\big\{(1-2\zeta\alpha\tilde{z})^2-\zeta^2\big\}
\tilde{\rho}^4
\big]
,\nonumber\\
F_z(\tilde{\rho},\tilde{z})=&\,
\frac{m \omega_z^2 \zpfz }
{\zeta}
\big[(
\sqrt{2}\xi
l_z\alpha-\tilde{z})\zeta
+2(1-2\zeta\alpha\tilde{z})\alpha\tilde{\rho}^2 
\big]
.
\end{align}
The set of the above relations and definitions 
represents a system of coupled ordinary differential equations;
their complexity inhibits the determination of exact analytical solutions. 
However the parameter $\alpha$ is often small, see Table~\ref{table:1} with the values suggested for the quantum logic scheme.
One ingredient for further progress 
therefore involves a perturbative treatment in $\alpha$.

The second ingredient consists of the linearization 
of the system~(\ref{rho_z_EoM}).
Many stable circular-orbit trajectories 
with constant $\tilde{\rho}=\tilde{\rho}_0$ and $\tilde{z}=\tilde{z}_0$ 
continue to satisfy the full equations of motion~(\ref{fullEoM}). 
More general solutions can then be constructed 
by considering small oscillations $\delta\tilde{\rho}$, $\delta\tilde{z}$ of the charge 
about $\tilde{\rho}_0$ and $\tilde{z}_0$. 
This idea can be realized 
via the ordinary method of expanding the effective forces 
$F_\rho(\tilde{\rho},\tilde{z})$ and $F_z(\tilde{\rho},\tilde{z})$
about $F_\rho(\tilde{\rho}_0,\tilde{z}_0)=0$ and $F_z(\tilde{\rho}_0,\tilde{z}_0)=0$ 
to linear order in $\delta\tilde{\rho}$ and $\delta \tilde{z}$.

Such a linearization 
provides perturbative solutions to the reduced system~(\ref{rho_z_EoM}), 
which can in turn be employed 
to determine the $\phi$ motion. 
The details of this analysis
are straightforward 
and have been relegated to Appendix~\ref{linear}.
The result completely characterizes 
the perturbative bound-state solutions 
to the equation of motion~(\ref{fullEoM})
at ${\cal O}(\alpha^2, a_\rho, a_z)$,
where $a_\rho$ and $a_z$ are the amplitudes defined in Appendix~\ref{linear}.
This solution may be presented 
in various other ways.
Here, 
we consider two variants
of expressing our solution
because each allows for distinct insights into 
the corresponding orbit 
and oscillation frequencies. 

\begin{figure}[t]
	\centering
	\includegraphics[width=\columnwidth]{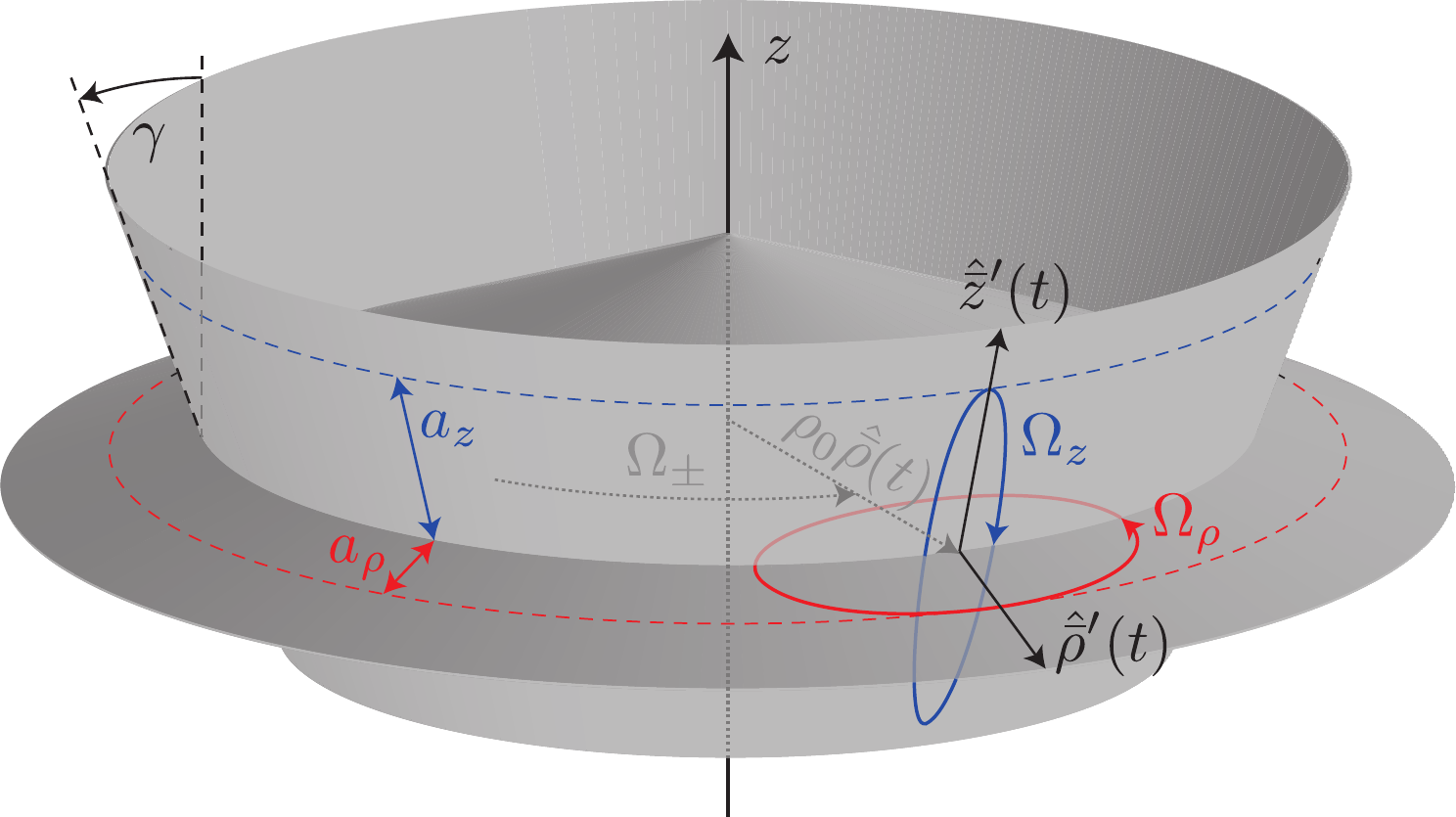}
	\caption{Motional modes 
	in a Penning trap 
	with nonzero $\vec{B}$-field gradient $b\neq 0$. 
	The mode shown in gray represents 
	circular motion about the $z$ axis; 
	it has served as the anchor for our perturbative treatment.
	The mode is described by 
	the vector $z_0\hat{z}+\rho_0\hat{\bar{\rho}}(t)$ 
	with $\hat{\bar{\rho}}(t)$ perpendicular to the $z$ axis.
	The orbit's radius $\rho_0$ 
	and its $z$ component $z_0$ 
	are determined by Eq.~(\ref{phys-roots}).
	The angular frequency of the rotation 
	of $\hat{\bar{\rho}}(t)$ about $\hat{z}$ 
	is $\Omega_+$ if $L_z>0$ 
	or $\Omega_-$ if $L_z<0$,
	as given by Eq.~(\ref{magnetron}).
	In this work, 
	the remaining two modes have been treated 
	as perturbations about this circular orbit 
	with small amplitudes $a_\rho$ and $a_z$.
	They are most easily described 
	relative to the rotating tip of the vector
	$z_0\hat{z}+\rho_0\hat{\bar{\rho}}(t)$. 
	Both modes are then small, 
	mutually orthogonal elliptical orbits 
	centered at this vector.
	The $a_\rho$ mode (red) and the $a_z$ mode (blue)
	are both tilted in the $\hat{\bar{\rho}}(t)$--$\hat{z}$ plane 
	by the angle $\gamma$ in Eq.~(\ref{modedirections})
	relative to the disk $z=z_0$ 
	and the cylinder $\rho=\rho_0$, 
	respectively. 
	This results in the two cones shown.
	The corresponding angular frequencies 
	are $\Omega_\rho$ and $\Omega_z$ 
	determined by Eq.~(\ref{Omega+}).
	}
	\label{fig:figure2}
\end{figure}

The first of these is obtained by
introducing the vector
\begin{equation}
\label{ave_rho}
\vec{\bar{\rho}}(t)=\rho_0 \hat{\bar{\rho}}(t) + z_0 \hat{z}\,,
\qquad
\hat{\bar{\rho}}(t)=
\left(
\begin{array}{c}
\cos \Omega_\pm t \\
\sin \Omega_\pm t \\
0 \\
\end{array}
\right)
\end{equation}
describing uniform motion 
on the circular path~(\ref{phys-roots}) that 
serves as the anchor 
for our above perturbative treatment. 
It may be viewed as the average position of the charge
about which small oscillations in all three dimensions occur.
To characterize these,
we define the following orthonormal
moving frame:
\begin{align}
\label{frame}
\hat{\bar{\rho}}'(t) 
= &\;\hat{\bar{\rho}}(t)\cos \gamma+\hat{z}\sin \gamma\,,\nonumber\\
\hat{\bar{z}}'(t)
= &\;\hat{z}\cos \gamma-\hat{\bar{\rho}}(t) \sin \gamma\,,
\nonumber\\
\hat{\bar{\phi}}'(t)
= &\;\hat{z}\times\hat{\bar{\rho}}(t) \,.
\end{align}
A leading-order expression 
for tilt angle $\gamma$ 
can be found in Appendix~\ref{linear}.
With respect to this basis, 
and again omitting phases,
the orbit can be expressed as
\begin{align}
\label{frame_sol}
\vec{r}(t)
= \vec{\bar{\rho}}&\;
{}+a_\rho\left[\hat{\bar{\rho}}'\cos \Omega_\rho t
-\xi\,(1-{\rm e}_\rho)\, \hat{\bar{\phi}}'\sin \Omega_\rho t \right] 
\nonumber\\
&\;
{}+a_z\left[\hat{\bar{z}}'\cos \Omega_z t
+(1-{\rm e}_z)\, \hat{\bar{\phi}}'\sin \Omega_z t \right], 
\end{align}
where the parameters ${\rm e}_\rho$ and ${\rm e}_z$ are given by:
\begin{align}
\label{eccentricities}
{\rm e}_\rho=&\;
\frac{2\, \omega_c l_z \omega_z^3 
(4 \omega_c^2 \omega_{\pm}+\omega_c \omega_z^2-12 \omega_{\pm} \omega_z^2)}{(\omega_c^2 -3\omega _z^2)^2 
(\omega_+\! -\omega_-)^3}\,\alpha^2\,,
\nonumber\\
{\rm e}_z=&\;1-
\frac{2\sqrt{l_z \omega_z} (3 \omega _z^2-2 \omega _c
   \omega _{\pm})}{(\omega _c^2-3 \omega _z^2)
   (\omega_+\! -\omega_-)^{1/2}}\,\alpha\,.
\end{align}
It thus becomes apparent that 
there are three independent modes 
in the bound orbit of the charge $q$,
as shown in Fig.~\ref{fig:figure2}.
One of these is the uniform circular motion 
with amplitude $\rho_0$ and angular frequency $\Omega_\pm$.
The second mode 
with amplitude $a_\rho$ 
and angular frequency $\Omega_\rho$
corresponds to an elliptical path 
centered at $\vec{\bar{\rho}}$
in the plane spanned by 
$\hat{\bar{\rho}}'$ and $\hat{\bar{\phi}}'$. 
The eccentricity of the ellipse is governed by ${\rm e}_\rho$; 
in the limit $\alpha \to 0$ of an ordinary Penning trap,
${\rm e}_\rho \to 0$,
and the ellipse becomes a circle.
The third mode
with amplitude $a_z$ 
and angular frequency $\Omega_z$
also represents an elliptical path 
centered at $\vec{\bar{\rho}}$,
but in the plane spanned by 
$\hat{\bar{z}}'$ and $\hat{\bar{\phi}}'$. 
The eccentricity of this second ellipse 
is determined by ${\rm e}_z$; 
in the conventional-case limit $\alpha \to 0$,
${\rm e}_\rho \to 1$,
and the ellipse becomes a line.

While suitable for characterizing the geometric shape
of the charge's trajectory,
Eq.~(\ref{frame_sol}) obscures the trap frequencies
because the time dependence 
is carried by both the basis vectors 
and the concomitant vector components.
An expression for  $\vec{r}(t)$ 
with a more transparent time dependence
suitable for exposing the trap frequencies 
can be obtained as follows. 
Let us denote 
any unit vector perpendicular to the trap's axis 
and rotating with a frequency $\omega$ 
relative to the trap by 
\begin{equation}
\label{varrhodef}
\hat{\varrho}(\omega)\equiv
\left(
\begin{array}{c}
\cos \omega t \\
\sin \omega t \\
0 \\
\end{array}
\right).
\end{equation}
With this notation, 
and dropping phases as before,
the trap solution $\vec{r}(t)$ 
can alternatively be expressed as
\begin{align}
\label{trap_sol}
\vec{r}(t)
= &\; \rho_0\,\hat{\varrho}(\Omega_\pm)
+\left[z_0+a_z \cos\gamma \cos\Omega_z t\right]\hat{z}
\nonumber\\
&
{}+\tfrac{1}{2}a_\rho\left[\cos\gamma+(1-{\rm e}_\rho)\xi\right]
\hat{\varrho}(\Omega_\pm-\Omega_\rho)
\nonumber\\
&
{}+\tfrac{1}{2}a_\rho\left[\cos\gamma-(1-{\rm e}_\rho)\xi\right]
\hat{\varrho}(\Omega_\pm+\Omega_\rho)
\nonumber\\
&
{}-\tfrac{1}{2}a_z\left[\sin\gamma+(1-{\rm e}_z)\right]\hat{\varrho}(\Omega_\pm-\Omega_z)
\nonumber\\
&
{}-\tfrac{1}{2}a_z\left[\sin\gamma-(1-{\rm e}_z)\right]\hat{\varrho}(\Omega_\pm+\Omega_z)
\nonumber\\
&{}+a_\rho \sin\gamma \cos\Omega_\rho t\,\hat{z}\,.
\end{align}
It is apparent that 
in this expression for $\vec{r}(t)$,
the time dependences
are separated in the desired fashion:
the modified axial frequency 
can be inferred from the coefficient multiplying $\hat{z}$,
and the radial frequencies 
are given as arguments of $\hat{\varrho}$.


To extract from Eq.~(\ref{trap_sol}) 
the corrections to the usual trap frequencies $\omega_z$ and $\omega_\pm$,
we disregard those modes 
whose amplitudes are suppressed by $\alpha$. 
The last three lines of Eq.~(\ref{trap_sol}) and, 
depending on the sign of $\xi$,
one of the $a_\rho$ modes 
can then be dropped.
It is thus apparent that
the modes with 
$\Omega_z$, $\Omega_\pm$, and $\Omega'_\mp\equiv\Omega_\pm \mp \Omega_\rho$ 
with
\begin{equation}
\label{3rdF}
\Omega'_\mp = \omega_\mp
\pm\frac{2 \omega_z^3 (\omega_c^2-6\omega_z^2)}{(\omega_c^2-3 \omega_z^2) (\omega_c^2-2 \omega_z^2)}
l_z \alpha^2
\end{equation}
survive 
and represent the generalizations 
of the usual trap frequencies.
In summary,
we find that
for $L_z>0$,
the usual trap frequencies 
are modified according to
$\{\omega_z,\omega_+,\omega_-\}\to\{\Omega_z,\Omega_+,\Omega'_-\}$ 
and for $L_z<0$
according to
$\{\omega_z,\omega_+,\omega_-\}\to\{\Omega_z,\Omega'_+,\Omega_-\}$. 
Experiments often involve situations
in which the mode with the smallest frequency 
will exhibit the largest amplitude.
Since our small-oscillation approach necessitates 
$\rho_0 \gg a_\rho$,
we conclude that 
the $\rho_0$ oscillations have smaller frequencies 
than $a_\rho$ oscillations. 
In this scenario, 
the second of the above assignments 
$\{\omega_z,\omega_+,\omega_-\}\to\{\Omega_z,\Omega'_+,\Omega_-\}$ 
is the relevant one.

Thus far, 
any effects due to the particle's spin $\vec{s}$ 
and the associated magnetic moment $\vec{\mu}=\frac{gq\hbar}{2m}\vec{s}$, 
where $g$ denotes the $g$ factor, 
have been disregarded. 
Although spin is best treated quantum mechanically 
as in Secs.~\ref{QTrap} and~\ref{NTrap}, 
our goal here is to develop some intuition 
about the corrections to the classical trajectory~(\ref{frame_sol}) 
or equivalently~(\ref{trap_sol}) 
due to the charge's intrinsic spin. 
To this end,
our dynamical system now 
consists of Eq.~(\ref{fullEoM}) 
supplemented by a magnetic-force term
$\vec{F}_d\equiv\vec{\nabla}\,(\vec{\mu}\cdot\vec{B})$. 
In what follows,
we consider spin alignments (anti)parallel to the magnetic field. 
In Appendix~\ref{spin} 
we argue that these two configurations are approximately static.
The magnetic force can then be expressed as
\begin{equation}
\label{Fmag}
\vec{F}_d=-\sigma\frac{|q|gs\hbar}{2m}\vec{\nabla}|\vec{B}|\,,
\end{equation} 
where $\sigma\equiv-\frac{q}{|q|s}(\vec{s}\cdot\hat{B})=\pm 1$ 
parametrizes our binary spin choices
with the positive (negative) sign 
corresponding to spin-up (spin-down) 
configurations relative to 
$-q\vec{B}$. 

Inclusion of the force~(\ref{Fmag}) 
into the trap equation of motion~(\ref{fullEoM}) 
yields perturbative solutions identical in structure 
to their spinless analogues~(\ref{frame_sol}) 
and~(\ref{trap_sol}) 
with definitions~(\ref{ave_rho}) and (\ref{frame}). 
However,
the expressions for the equilibrium coordinates $\rho_0$ and $z_0$
as well as the frequencies $\Omega_\rho$ and $\Omega_\pm$ 
appearing in these equations 
acquire spin-dependent corrections:
\begin{align}
\label{spin_corrections}
&\tilde{\rho}_0=\;
\frac{2^{1/4}\sqrt{l_z\zeta}}{(1-\zeta^2)^{1/4}}\nonumber\\
&\;{}+\frac{2^{3/4}l_z(1-\zeta^2)^{-1/2}+\xi\, l_z
+\sigma gs(1-\tfrac{1}{4}\zeta^2)}
{(1-\zeta^2)^{5/4}}\zeta\sqrt{l_z\zeta}\alpha^2\,,
\nonumber\\
&\tilde{z}_0=\;
\left(
\xi+\frac{1}{\sqrt{1-\zeta^2}}
\right)\sqrt{2} l_z\alpha+ \sqrt{2} \sigma g s\,\alpha\,,\nonumber\\
&\Omega_\rho=\;\sqrt{\omega_c^2-2\omega_z^2}
+ 4 \omega_z\left[\vphantom{\frac{\xi\omega_c^2 l_z+\sigma g s(\omega_c^2-\frac{1}{2}\omega_z^2)}{\omega_c\sqrt{\omega_c^2-2\omega_z^2}}}
\frac{\omega_c^2\omega_z^2 l_z}{2(\omega_c^2-3\omega_z^2)(\omega_c^2-2\omega_z^2)}\right.
\nonumber\\
&\;{}
-\left.
\frac{\omega_c^2 l_z}{\omega_c^2-2\omega_z^2}
-\frac{\xi\omega_c^2 l_z+\sigma g s(\omega_c^2-\frac{1}{2}\omega_z^2)}{\omega_c\sqrt{\omega_c^2-2\omega_z^2}}
\right]\alpha^2\,,\nonumber\\
&\Omega_\pm=\;
\omega_\pm
+4\, \frac{\omega_z (\omega _z^2-2 \omega_c \omega_{\pm})}{\omega_c^2-2 \omega_z^2}\,\xi\,l_z\alpha^2
\nonumber\\
&\;{}-2\,\sigma g s\,  \omega_z  \left(1+\xi\,
\frac{2 \omega_c^2-\omega_z^2}{2 \omega_c \sqrt{\omega_c^2-2 \omega_z^2}}\right)\alpha^2\,.
\end{align}
We note that the eccentricities~(\ref{eccentricities}) remain spin independent.

The above analysis demonstrates that
the inclusion of a linear magnetic-field gradient 
into a conventional Penning trap 
leads to several modifications 
in the trapped charge's classical orbit.
The axial equilibrium position 
and the orientation of the normal modes 
acquire corrections at linear order 
in the $\vec{B}$-field gradient, 
which are given in Eqs.~(\ref{spin_corrections}) and~(\ref{mode_angle_appr}),
respectively.
Frequency mixing between the various degrees of freedom 
deforms the usual linear $\omega_z$ mode 
and one of the circular modes 
(i.e., the one typically associated with $\omega_m$) 
into ellipses
with eccentricities determined by Eq.~(\ref{eccentricities}).
These corrections are of linear and quadratic order 
in the $\vec{B}$-field gradient,
respectively.
The conventional trap frequencies 
are modified only at second order in the gradient;
the perturbative expressions for them 
are regime dependent 
and follow from Eqs.~(\ref{Omega+}) and~(\ref{spin_corrections}),
as explained in the context of Eq.~(\ref{3rdF}). 
The orientation of the charge's spin 
affects only the subset of classical-orbit parameters 
displayed in Eq.~(\ref{spin_corrections}).

\section{Conclusions}
In this article we have studied an \ppbar in a Penning trap with a superimposed magnetic field gradient. We have shown that the magnetic field gradient allows the implementation of elementary laser-less quantum logic operations, in particular of a SWAP gate between the spin and axial motion degrees of freedom, as required for the realization of quantum logic spectroscopy in this system. Through numerical simulations, we predict that error probabilities on the per mil level for viable trap and pulse parameters are achievable. We give an intuitive classical picture of the motion in the Penning trap with a strong superimposed magnetic field gradient.

\begin{acknowledgments}
This work was supported by the DFG through CRC 1227 `DQ-mat' projects A06 and B06, the cluster of excellence `Quantum Frontiers', ERC StG `QLEDS' and the Indiana University Center for Spacetime Symmetries. R.L.~acknowledges support from the Alexander von Humboldt Foundation.
\end{acknowledgments}

\bibliography{bibliography}


\begin{appendix}
\begin{widetext}
\section{Discussion $H_2$}\label{sec:dis}
The full Hamiltonian including all terms from the transverse oscillating field is (with $2\Omega_c\approx\omega_c$)
\begin{align}&H_2=-\frac{\Omega}{g}\frac{\hbar}{2}\sqrt{\frac{\omega_c}{\omega_z}}(a_z+a_z^\dagger)(e^{i\omega t}a_c^\dagger+e^{-i\omega t}a_c)\nonumber\\
&-\frac{\Omega}{g}\frac{\hbar}{2}\sqrt{\frac{\omega_z}{\omega_c}}(e^{i\omega t}(a_m+a_c^\dagger)+e^{-i\omega t}(-a_c-a_m^\dagger))(a_z-a_z^\dagger)\nonumber\\
&-\frac{\Omega}{g}\frac{\hbar}{2}\sqrt{\frac{\omega_c}{\omega_z}}\epsilon(a_z+a_z^\dagger)^2(e^{i\omega t}(a_m+a_c^\dagger)+e^{-i\omega t}(a_c+a_m^\dagger))\nonumber\\
&+\frac{\Omega}{g^2}\frac{\hbar}{4}\frac{\Omega}{\omega_z}(a_z+a_z^\dagger)^2+\frac{\Omega}{g^2}\frac{\hbar}{2}\frac{\Omega}{\omega_c}(a_ca_m+a_c^\dagger a_m^\dagger +a_c^\dagger a_c +a_m^\dagger a_m +1)\nonumber\\
&+\frac{\Omega}{g^2}\frac{\hbar}{4}\frac{\Omega}{\omega_c}e^{2i\omega t}(-a_c^\dagger a_c^\dagger-a_ma_m-2a_ma_c^\dagger)\nonumber\\
&+\frac{\Omega}{g^2}\frac{\hbar}{4}\frac{\Omega}{\omega_c}e^{-2i\omega t}(-a_ca_c-a_m^\dagger a_m^\dagger-2a_ca_m^\dagger)+\Omega\frac{\hbar}{2}(\sigma^+e^{-i\omega t}+\sigma^-e^{i\omega t}). \nonumber
\end{align}
Considering the parameters presented in Table~\ref{table:1} of the main text we find that only the first term on the right hand side is of similar magnitude compared to the resonant sideband driving, resulting from the last term. We can however also disregard the first term as all its contributions will be off-resonant with the cyclotron frequency $\omega_c$ when going to an interaction picture with respect to all motional modes, similar to Sec.~\ref{sec:Title3}. With $\omega_c \gg \omega_z$ the effect of these terms will also be suppressed.

\end{widetext}

\section{Linearization}\label{linear}

The implementation of the mode linearization 
yields 
\begin{align}
\label{phys-roots}
\tilde{\rho}_0=&\;
\frac{2^{1/4}\sqrt{l_z\zeta}}{(1-\zeta^2)^{1/4}}
+\frac{2^{3/4}(l_z\zeta)^{3/2}(1+\xi\sqrt{1-\zeta^2})}
{(1-\zeta^2)^{7/4}}\alpha^2\,,
\nonumber\\
\tilde{z}_0=&\;
\left(
\xi+\frac{1}{\sqrt{1-\zeta^2}}
\right) \sqrt{2} l_z\alpha
\end{align}
at leading order in $\alpha$ 
for the equilibrium circular orbits. 
These solutions represent perturbations 
relative to the conventional Penning trap.
Note that $\tilde{\rho}_0 \to 0$ 
for $l_z \to 0$, 
so that this limit inhibits a linearization approximation 
since the size $\delta\tilde{\rho}<\tilde{\rho}_0$ of small harmonic oscillations  
is effectively squeezed to zero.
Note also that 
additional solutions exist
that may in principle be physical. 
It might be interesting to investigate their stability, 
but this lies outside our present scope.

The eigenfrequencies
of the system~(\ref{rho_z_EoM}) 
about the equilibrium~(\ref{phys-roots})
can then be determined as usual:
\begin{align}
\label{Omega+}
\Omega_\rho=&\,
\sqrt{\omega _c^2-2 \omega_z^2}\nonumber\\
&\,{}-\left[
\frac{4 \xi  \omega_c \omega_z}{\sqrt{\omega_c^2-2 \omega_z^2}}
+\frac{2\omega_c^2 \omega_z (2 \omega_c^2-7\omega_z^2)}{(\omega_c^2-3 \omega_z^2) (\omega_c^2-2 \omega_z^2)}
\right]l_z \alpha^2  
\,,
\nonumber\\
\Omega_z
=&\;
\omega_z
-\frac{6 \omega_z^4}
{(\omega_c^2-3\omega_z^2) \sqrt{\omega_c^2-2 \omega_z^2}}\, 
l_z \alpha^2\,.
\end{align}
This form of the eigenfrequencies,
valid at leading order in $\alpha$,
has been derived under the assumption that 
$(2-3\zeta^2)>0$ is not too close to zero.
The normal-mode directions 
$\hat{\rho}'$ and $\hat{z}'$ 
associated to the respective frequencies 
$\Omega_\rho$ and $\Omega_z$
are tilted relative to the conventional modes
along $\hat{\rho}$ and $\hat{z}$ 
by an angle $\gamma$
\begin{align}
\label{modedirections}
\hat{\rho}' 
= &\;\hat{\rho}\cos \gamma-\hat{z}\sin \gamma\,,\nonumber\\
\hat{z}' 
= &\;\hat{z}\cos \gamma+\hat{\rho} \sin \gamma\,.
\end{align}
where
\begin{equation}
\label{mode_angle_appr}
\gamma=\frac{2^{5/4}\zeta^{3/2}}{(2-3\zeta^2)(1-\zeta^2)^{1/4}}\,
\sqrt{l_z}\,\alpha\,,
\end{equation}
for small $\alpha$ and $(2-3\zeta^2)>0$. 
Together with the amplitudes and phases for these two modes 
representing four integration constants,
these results provide a full description of the system~(\ref{rho_z_EoM}).

The remaining task is 
to extract the angular motion $\phi(t)$.
This can, 
for example, 
be achieved with the above results and Eq.~(\ref{Lz}):
\begin{equation}
\label{phi_sol}
\phi(t)
=\Omega_\pm t
+c_{\rho}\sin\Omega_{\rho} t 
+c_z\sin\Omega_z t\,,
\end{equation}
where
\begin{equation}
\label{magnetron}
\Omega_\pm=
\omega_\pm
+4\, \frac{\omega_z (\omega _z^2-2 \omega_c \omega_{\pm})}{\omega_c^2-2 \omega_z^2}\,\xi\,l_z\alpha^2\,.
\end{equation}
The coefficients $c_\rho$ and $c_z$ 
are uniquely determined by the amplitudes $a_\rho$ and $a_z$ 
as well as other system parameters;
their determination is straightforward 
but the expressions are not particularly illuminating.
The $\pm$ sign choices in Eq.~(\ref{magnetron}) 
are correlated with $\xi=\pm 1$.
For brevity,
we have omitted the aforementioned $\Omega_\rho$ and $\Omega_z$ phases 
as well as a new phase associated with $\Omega_\pm$.

\section{Spin motion}\label{spin}

The nonrelativistic precession of spin is governed by the equation 
\begin{equation}
\label{SpinEoM}
\dot{\vec{s}}=\frac{gq}{2m}\vec{s}\times\vec{B}(\vec{r})\,.
\end{equation} 
We note in passing that 
this equation implies the conservation of $|\vec{s}|\equiv s$
paralleling the quantum description.

The initial spinless analysis in Sec.~\ref{classical} 
was performed pertubatively in $\alpha \propto b$, 
suggesting an analogous approach for $\vec{s}$.
We are thus led to decompose the spin motion $\vec{s}(t)$ 
as 
\begin{equation}
\label{spin_ansatz}
\vec{s}(t)=\vec{s}_0(t)+\delta\vec{s}(t)\,,
\end{equation}
where $\vec{s}_0(t)=s_\parallel\hat{B}+s_\perp\hat{e}_\perp(t)$ 
describes spin precession 
about the local $\vec{B}$-field direction
$\hat{B}(\vec{r})\equiv\vec{B}(\vec{r})/|\vec{B}(\vec{r})|$,
and 
$\delta\vec{s}(t)=\delta s_\parallel(t)\hat{B}+\delta s_\perp(t)\hat{e}'_\perp(t)$
a correction at most of ${\cal O}(b)$.
Here, 
$\hat{e}_\perp(t)$ and $\hat{e}'_\perp(t)$ 
are unit vectors perpendicular to the local $\vec{B}(\vec{r})$ field.
We also select $\dot{s}_\parallel=0$.

With these considerations, 
we may gain insight 
into the time evolution of the spin projection $\vec{s}\cdot\hat{B}=s_\parallel+\delta s_\parallel$.
As Eq.~(\ref{SpinEoM}) 
implies $\dot{\vec{s}}\cdot\hat{B}=0$,
we obtain 
\begin{equation}
\label{time_der2}
\delta\dot{s}_\parallel=
\vec{s}\cdot\dot{\hat{B}}
=
\big[s_\perp\hat{e}_\perp(t)+\delta s_\perp\hat{e}'_\perp(t))\big]\cdot\dot{\hat{B}}\,.
\end{equation}
If we now specialize to perturbations $\delta\vec{s}(t)$ 
about spin alignments $\vec{s}(t)$ 
along or opposite to the local $\vec{B}(\vec{r})$ field,
i.e., 
$s_\perp(t)=0$, 
we have
\begin{equation}
\label{time_der3}
\delta\dot{s}_\parallel=
{\cal O}(b^2)\,,
\end{equation}
since $\dot{\hat{B}}={\cal O}(b)$.
This shows that spin perturbations 
away from an initial parallel or antiparallel configuration
are suppressed. 
It is therefore justified 
to take these two configurations 
as approximately static,
as advertised in the main text.

\end{appendix}




\end{document}